# LABORATORY OPTICAL SPECTROSCOPY OF
# VIBRONIC TRANSITIONS OF THE THIOPHENOXY RADICAL

Haruka T. Sato,[1] Mitsunori Araki,[1,2] Takahiro Oyama,[1] and Koichi Tsukiyama[1]

[1] Department of Chemistry, Faculty of Science Division I, Tokyo University of Science,
1-3 Kagurazaka, Shinjuku-ku 162-8601, Tokyo, Japan; araki@rs.tus.ac.jp
[2] Research Institute for Science and Technology, Tokyo University of Science,
2641, Yamazaki, Noda, Chiba, 278-8510, Japan

## ABSTRACT

The thiophenoxy radical ($C_6H_5S$) is a species of possible astrophysical interest due to an electronic transition in a 5000 Å region. The $B\,^2A_2 \leftarrow X\,^2B_1$ electronic transition of this radical in the discharge of thiophenol was measured using a cavity ring-down spectrometer. The optical absorption spectrum of this transition was obtained in the range covering from the origin band (0–0) to a frequency of 1750 $cm^{-1}$. The vibronic bands in the 400–1700 $cm^{-1}$ region are stronger than the origin band, suggesting structural difference between the ground and excited electronic states. The prominent progression was assigned to the 6a symmetric in-plane CCC bending mode starting from the $6b_0^1$ forbidden band. Band origins of individual bands were determined by analysis of the rotational profiles. Although these vibronic bands were not found in optical spectra of diffuse clouds, the upper limits of the column densities for the thiophenoxy radical in the diffuse clouds toward HD 183143 and HD 204827 were evaluated to be ~$4 \times 10^{13}$ $cm^{-2}$.

*Subject Keywords*: Astrochemistry—ISM: clouds—ISM: molecules—ISM: lines and bands

## 1. Introduction

The chemical composition of the universe is a topic that has occupied scientists for centuries. Absorption bands of molecules in diffuse cloud in the visible and near-infrared regions have been detected as diffuse interstellar bands (DIBs) since 1919 (Heger 1922, McCall & Griffin 2013). Identifications of DIBs are means to reveal the chemical composition. A variety of molecules are thought to exist in diffuse clouds. Identifications of DIB carrier materials are one of the longest-standing unsolved problems in spectroscopy and astrochemistry (Herbig 1995, Cami & Cox 2014). Identifications can be carried out, firstly, by measuring optical spectra of DIB carrier candidate molecules in laboratory and, subsequently, by comparing laboratory spectra with astronomically observed DIB spectra. Thus, prior to the identification of the potential carrier involved in DIBs, the spectral characteristics of candidate molecules must be studied in detail in the laboratory. The first identification of DIBs was achieved in 2015 via laboratory work by Campbell et al. (2015). The five DIBs in all were assigned to the electronic transition of $C_{60}^+$ (Walker et al., 2015, Campbell et al., 2016a, 2016b). Prior to this discovery, the size of the largest interstellar molecules undoubtedly identified consisted of 12 atoms, e.g., trans-ethyl methyl ether $C_2H_5OCH_3$ (Fuchs et al. 2005) and so on, while benzonitrile consisting of 13 atoms was found by McGuire et al. (2018) after the assignment of $C_{60}^+$. Existence of $C_{60}^+$ drastically extended size variation of interstellar molecules, which raises an issue for production scheme of molecules in space. Generation of almost all interstellar molecules has been explained by bottom up chemistry, in which larger molecules are produced from smaller molecules via ion-neutral, neutral-neutral and grain surface reactions. However, this identification of $C_{60}^+$ suggests a possibility of top down chemistry, in which molecules are obtained from dissociations and/or deformations of larger molecules (e.g., Linnartz et al. 2020, Oka & Witt 2016). Thus, searches for large interstellar molecules may give us key information to infer the production scheme of molecules.

We presume that the thiophenoxy radical ($C_6H_5S$), a typical sulfur-atom substituted small aromatic compound, is a prevailing candidate for DIB carrier because more than 10 % of interstellar molecular species include a sulfur atom: in fact, sulfur-atom substituted $C_5$ linear carbon chain, $C_5S$, is comparably abundant with hydrogen-atom substituted species, $C_5H$, in a circumstellar envelope (Agúndez et al. 2014, Cernicharo et al. 1986). Even if ionized species is more abundant than neutral species in diffuse clouds, the thiophenoxy radical would still exist, because this radical can be produced through dissociation of the thiophenol cation (e.g., Hermann et al. 2000). The small aromatic compounds tend to exhibit electronic absorption bands in the UV region, while their radical species generally have optical electronic transitions in the visible region. Substituted aromatic compounds generally have larger permanent dipole moments than non-substituted ones and suffer radiative cooling of their rotational temperatures in space. Its narrow rotational distribution at low temperature, giving rise to strong peaks in the electronic transitions because of piling up of rotational lines, is advantageous for detecting the thiophenoxy radical in space.

The laboratory absorption spectrum of the thiophenoxy radical in the 450–520 nm region, recorded with a photoplate, was reported by Okumura et al. (1977) without any description of line intensities. They suggested small structural difference between the excited and ground electronic states because no distinct vibrational progression was observed. Using laser induced fluorescence spectroscopy, the spectrum in the same wavelength region was studied by Shibuya et al. (1988). The origin (0 – 0 band) and a low frequency vibronic band at 504.76 nm were strongly observed, where the resolution of 1 Å was still insufficient to assign DIBs. Taking the band intensities and laser power into consideration, it was supposed that the strongest vibronic band in space is the origin band. Since a stellar atmospheric line exists at 5048 Å, to date (Hobbs et al. 2008, 2009), only the origin band allows us to compare laboratory data with DIBs. Based on this circumstance, the origin band was measured using high-resolution spectrometer in laboratory by Araki et al. (2014), hereafter referred as Paper I. They assigned the electronic transition to be $B\,^2A_2 \leftarrow X\,^2B_1$ and derived the upper limit of the column density in space from the comparison of the origin band with DIBs.

The phenoxy radical ($C_6H_5O$) has an isoelectronic structure with the thiophenoxy radical for valence electrons. The $B\,^2A_2 \leftarrow X\,^2B_1$ electronic transition of $C_6H_5O$ was investigated by Araki et al. (2015), and the upper limit of column density in a diffuse cloud was derived to be large ($7 \times 10^{14}$ cm$^{-2}$) because of lifetime broadening of lines. The vibronic bands were stronger than the origin band, and the long progression of the 6a symmetric in-plane CCC bending mode appeared. Accordingly, a long progression of the 6a mode can be also expected for the thiophenoxy radical. Indeed, the absorption spectrum of the thiophenoxy radical reported by Okumura et al. (1977) has shown signs of higher frequency vibronic bands even though they are unclear. The spectral characterization of such stronger vibronic bands at laboratory would enable us to explore this radical in space in much wider wavelength range.

In this work, the absorption spectrum of the $B\,^2A_2 \leftarrow X\,^2B_1$ electronic transition of the thiophenoxy radical was precisely measured by cavity ring-down (CRD) spectroscopy. We make unambiguous assignment of vibronic bands on the bases of their rotational profiles. Finally, by estimating astronomically expected rotational profiles of these bands, we discuss the abundance in space.

## 2. Experimental Details

A CRD spectrometer was used to observe the optical absorption spectrum in the wavelength range of 4722–5190 Å, which covers all absorption lines reported by Okuyama et al. (1977). This laboratory spectrometer has been described in detail elsewhere (Paper I). The 10-Hz tunable pulsed laser beam was taken from a dye laser (FL3002, Lambda physics) with a resolution of 0.2 cm$^{-1}$ and pumped by a Nd:YAG laser (355 nm, Surelit). The optical cavity was constructed with two high-reflectivity mirrors ($R > 99.99\%$, Los Gatos Research). To cover the wavelength range of 4722–5190 Å, two pairs of mirrors with different central wavelengths were used. The discharge to produce the thiophenoxy radical was created by a pulsed voltage with a pulse width of 1 ms starting ~700 μs ahead of a laser pulse. The cathode with an inner diameter of 15 mm and length of 100 mm made of stainless steel was newly installed in the glow-discharge cell illustrated in Figure 1 of Paper I. The thiophenoxy radical was produced by the discharge of 800 V with gas sample mixture of thiophenol ($C_6H_5SH$, 0.01 Torr) and helium (0.2 Torr). The discharge cell was cooled by liquid nitrogen to remove hot bands. The obtained spectrum was calibrated by using the wavelength meter SHR (SOLAR Laser Systems) in the accuracy of 0.03 Å.

## 3. Results and Discussion
### 3.1. *Vibrational Assignments*

The observed absorption spectrum of the thiophenoxy radical is shown in Figure 1. A number of peaks including the strongest one at 1281.2 cm$^{-1}$ appear in the higher frequency region of 800–1700 cm$^{-1}$ as listed in Table 1. Unlike the previous suggestion by Okuyama et al. (1977), this radical definitely has a large structural difference between the excited and ground electronic states. In Paper I, it was assumed that almost of all intensity of this electronic transition is distributed to the origin band. However, the present spectrum shows that the intensity distribution to the origin band is only 1.5 %. Thus, the upper limit derived in Paper I needs to be re-estimated.

First, we discuss the vibrational structure of this radical. The low-frequency peak appearing at 405.5 cm$^{-1}$ makes a progression together with the peak at 815.9 cm$^{-1}$. These peaks exhibit b-type rotational profiles, which are expected for allowed vibronic transitions of an $A_2 - B_1$ electronic transition. Theoretical calculations by B3LYP/6-311++G(d,p) predict the 6a CCC in-plane symmetric bending mode of 401 cm$^{-1}$ in the $B\,^2A_2$ excited state (Table A1). Franck-Condon calculation derives the strong progression of the 6a mode as shown in Figure A1. Thus, the progression starting from 405.5 cm$^{-1}$ can be safely assigned to the 6a mode. The first peak apparently corresponds with the band at 410 cm$^{-1}$ reported by Shibuya et al. (1988). Although this band was reported as of the 7a symmetric CS stretching mode, the frequency of this mode was calculated to be 1443 cm$^{-1}$ in the present work.

The strongest progression starts from the peak at 4

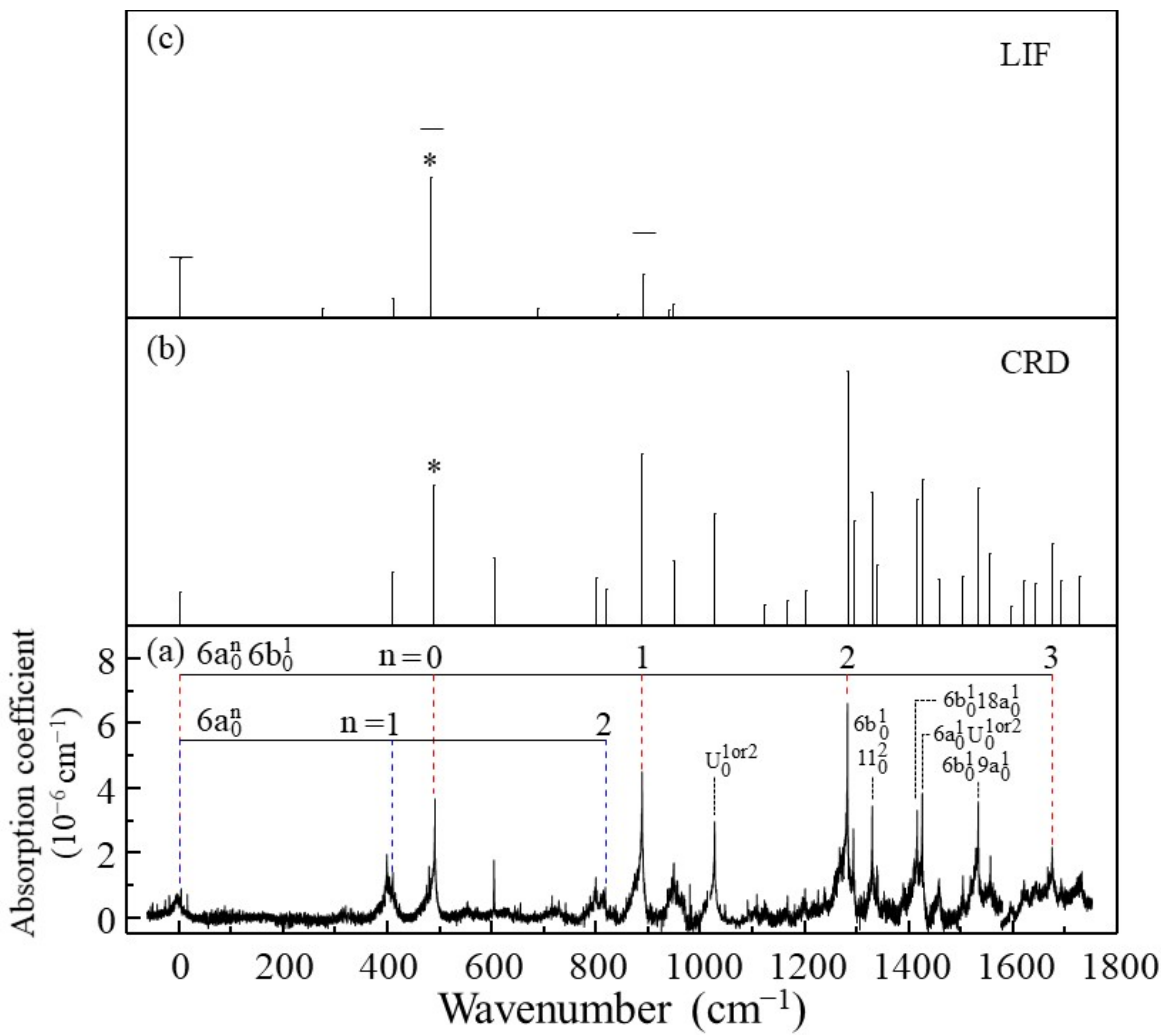

**Figure 1.** Observed absorption spectrum of the $B\,^2A_2 \leftarrow X\,^2B_1$ electronic transition of the thiophenoxy radical by cavity ringdown (CRD) spectroscopy (a). The 0 and 1600 cm$^{-1}$ positions correspond with 5172.5 and 4777.0 Å, respectively. (b) Stick diagram of the observed bands in the trace (a). (c) Stick diagram of the bands observed by laser induce florescence (LIF) spectroscopy (Shibuya et al. 1988). To compare the trace (c) with the trace (b), intensities in the trace (c) are normalized by that of the line marked by an asterisk. The crossbars over the three sticks in the trace (c) show relative intensities of laser.

87.7 cm$^{-1}$ with intervals of 393–399 cm$^{-1}$. All peaks of the progression have a-type rotational profiles. In the case of an $A_2 - B_1$ electronic transition, a-type transitions are classified as forbidden transitions because they are produced via intensity borrowing by vibronic interaction. The first peak can correspond to that at 483 cm$^{-1}$ (504.76 nm) reported by Shibuya et al. (1988). Although this peak was assigned to the $6a_0^1$ transition in their paper, the a-type rotational profile observed in the present work contradicts their assignment. To produce an a-type transition, a vibrational level in the excited state needs to be in b$_2$ symmetry in this electronic transition. Theoretical calculations give the 6b CCC in-plane anti-symmetric bending of 575 cm$^{-1}$ with b$_2$ symmetry (Table

A1). Aside from the 6b mode, no modes have acceptable frequencies. Although the 11 mode in Wilson' notation has the frequency of 397 cm$^{-1}$, the $11_0^1$ transition is symmetry-forbidden due to the b$_1$ symmetry of this mode. In this case, the transition is not allowed by electric dipole selection rules. Thus, the first peak at 487.7 cm$^{-1}$ can be assigned to the $6b_0^1$ forbidden transition. The intervals of 393–399 cm$^{-1}$ correspond to the 6a mode, and then this progression can be assigned to the combination bands of the 6a and 6b modes. The assigned vibronic bands are listed in Table 1.

For the thiophenoxy radical, the $C\,^2B_1$ electronic state lies 0.55 eV above the $B\,^2A_2$ electronic state (Paper I). The benzyl radical (C$_6$H$_5$CH$_2$), which has a comparable electronic structure with the thiophenoxy radical, also has the $^2B_1$ electronic state 0.22 eV above the $^2A_2$ electronic state (Tonokura & Koshi 2003). In the spectrum of the $^2A_2 \leftarrow X\,^2B_1$ electronic transition of the benzyl radical, the $6b_0^1$ forbidden vibronic bands get the strongest transition intensities by vibronic interaction of $^2A_2$ with $^2B_1$ through the 6b mode (Fukushima & Obi 1990, 1992). The $6b_0^1$ bands were observed to be sufficiently stronger than the origin band. Thus, because of similarity for the thiophenoxy radical, vibronic transitions involving the 6b mode originally with forbidden character are thought to obtain significant transition probabilities from the $C\,^2B_1 - X\,^2B_1$ electronic transition due to vibronic interaction of $B\,^2A_2$ with $C\,^2B_1$. Additionally, the strong appearance of the $6a_0^n\,6b_0^1$ progression is explainable by the large transition dipole moment of $f = 0.057$ (TD-B3LYP/cc-pVTZ) for the $C\,^2B_1 - X\,^2B_1$ electronic transition.

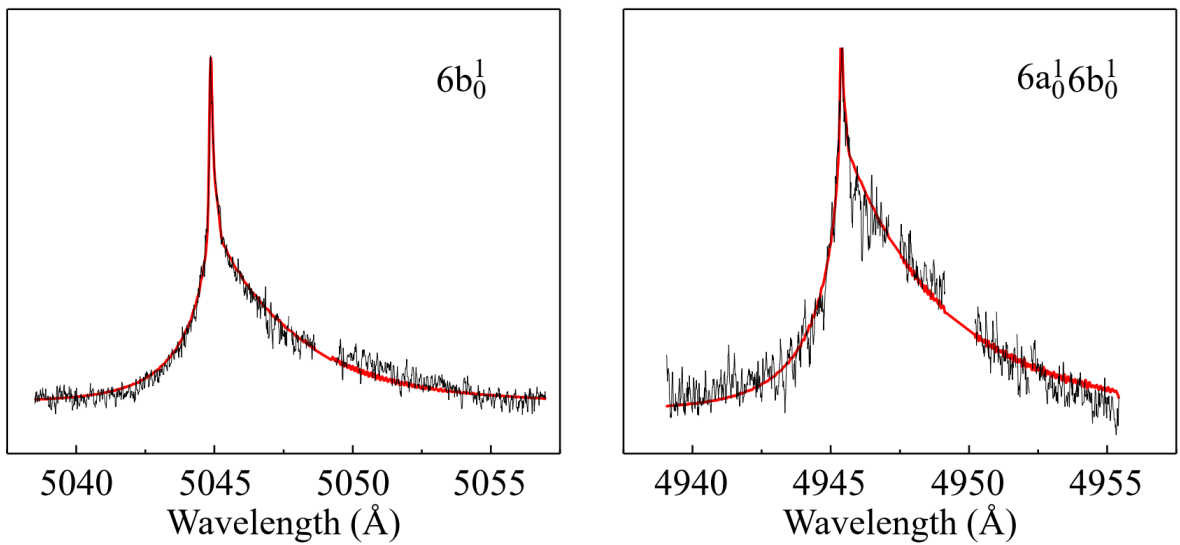

**Figure 2**. Observed (black trace) and simulated (red trace) rotational profiles of the $6b_0^1$ and $6a_0^1 6b_0^1$ vibronic bands in the $B\,^2A_2 - X\,^2B_2$ electronic transition of the thiophenoxy radical. PGOPHER (version 10.1.182, Western) was employed for simulation. Lacking parts in the traces of observed spectra are regions disturbed by spike noise due to discharge. These regions were excluded as data for profile fitting.

There are several absorption bands above 1000 cm$^{-1}$ with considerable intensities. All these peaks exhibit a-type profiles. The relatively strong peak at 1025.7 cm$^{-1}$ (hereafter $U_0^{1or2}$) cannot be assigned to a unique vibrational transition, e.g., $6b_0^1 18b_0^1$ is unlikely due to little Franck-Condon factor of $18b_0^2$. However, it is sure that this peak forms a progression of the 6a mode together with the peak at 1424.6 cm$^{-1}$. Aside from the 6a mode, the vibrations with $a_1$ symmetry combining with the 6b mode can produce a-type profiles. The peaks at 1327.8, 1414.0, and 1531.5 cm$^{-1}$ may be attributed to be $6b_0^1 11_0^2$, $6b_0^1 18a_0^1$, and $6b_0^1 9a_0^1$, respectively, based on frequencies and symmetries; however, these assignments are still tentative due to their higher frequencies.

### 3.2. *Rotational Analysis*

To compare the laboratory spectrum of the thiophenoxy radical with DIBs, rotational profiles in space need to be simulated. Observed profiles of the vibronic bands were analyzed to determine rotational constants by using PGOPHER (version 10.1.182, Western). Here we assume a rotational temperature of the thiophenoxy radical in the present experimental set up to be 340 K from the fitting of a profile of the origin band using the reported rotational constants (Paper I). This temperature was used to analyze profiles of the $6b_0^1$ and $6a_0^1 6b_0^1$ vibronic bands, which have no overlap with other bands. For the spectral analysis, all rotational constants in the ground state and $B - C$ in the excited state were fixed and a line width of each rotational line was assumed to be 0.08 Å as a summation of the laser linewidth (0.2 cm$^{-1}$) and an additional effect (0.1 cm$^{-1}$) from Doppler and/or lifetime broadening. By reproducing the observed profiles as shown in Figure 2, the band origins $T_{00}$ and the rotational constant differences $\Delta A$ and $\overline{\Delta B}$ were determined, where the band origin means the position of a rovibronic transition between $N = 0$ rotational revels. The determined constants are listed in Table 2. Except for the origin, $6b_0^1$, and $6a_0^1 6b_0^1$ bands, the band origins of vibronic bands were evaluated by profile fitting while all rotational constants were fixed (see Table 1).

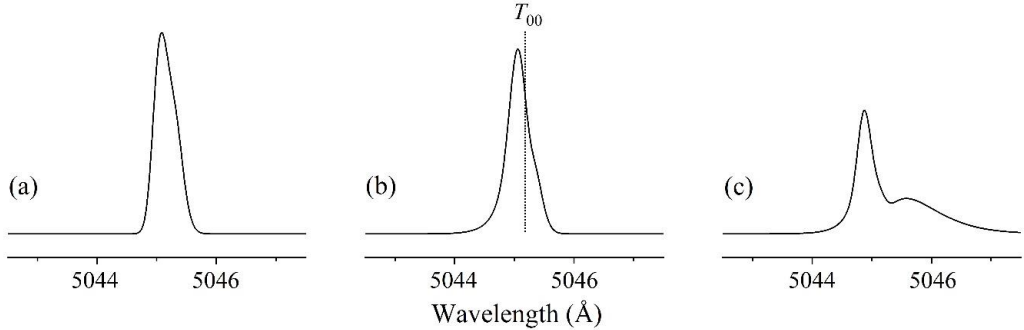

**Figure 3**. Simulated rotational profiles of the $6b_0^1$ vibronic band in the $B\,^2A_2 - X\,^2B_2$ electronic transition of the thiophenoxy radical. The simple Boltzmann distributions at $T_k = T_r = 2.73$ K (a) and 40 K (c), and the distribution at $T_k = 40$ K and $T_r = 2.73$ K in a diffuse cloud (b) are assumed. $T_{00}$ in the trace (b) indicates the band origin.

### 3.3. *Simulation of Rotational Profiles in Diffuse Clouds*

In addition to the rotational constants, rotational distributions are necessary to simulate absorption profiles in space. Since diffuse clouds are non-thermal equilibrium regions because of low number densities, a simple model assuming a Boltzmann distribution cannot be applied. For a near-prolate asymmetric top molecule such as the thiophenoxy radical, spontaneous emission between rotational levels $J$ and $J-1$ in the same $K_a$ has a large transition probability because of a permanent dipole moment along the molecular axis as shown in Figure 3 of Paper I, while that between different $K_a$ levels is very weak. The rotational distribution is determined by the balance between $J$ and $J-1$ under the coexistence of radiation and collisions, which are influenced by the radiative temperature $T_r$ and the kinetic temperature $T_k$. Rotation about the molecular axis could keep a higher temperature close to $T_k$, whereas overall rotation for $J$ is cooled to around $T_r$. This behavior can be described in equation (7) of Paper I as an effect of "hot axis." A rotational profile of the origin band with a b-type transition in the thiophenoxy radical was simulated in Paper I.

Rotational profiles of a-type transitions can be simulated by using the rotational constants determined in the previous section and rotational distribution derived by equation (7). As the excitation temperature of a molecule with no permanent dipole moment is equivalent with the kinetic temperature, we used $T_k = 40$ K. This temperature was estimated from the rotational structure of the $C_2$ spectrum toward HD 204827 (Oka et al. 2003). The collision rate was assumed to be $C = 10^{-7}$ s$^{-1}$ as a typical value of diffuse clouds (Oka & Epp 2004;

Oka et al. 2013). A resolution depending on the telescope and Doppler width of a cloud was set to be 0.22 Å as in Paper I. The rotational profile of the $6b_0^1$ vibronic band was simulated as the trace (b) in Figure 3. As comparison, rotational profiles assuming the simple Boltzmann distributions of 2.7 K (the cosmic background temperature) and 40 K were also illustrated in the traces (a) and (c), respectively.

The profile in the trace (b) is very similar to that of the trace (a), because rotational structures having different $K_a$ are almost overlapped due to small $\Delta A$. Then, a profile of an a-type transition has little dependence on a kinetic temperature of a cloud. Additionally, all the profiles of a-type vibronic transitions recorded in laboratory are nearly identical each other. This is because variations of $\Delta A$ and $\Delta \overline{B}$ depending on vibronic transitions are much smaller than $A$ and $\overline{B}$, respectively. Thus, under a low-temperature condition in space, the simulated profile in Figure 3(b) can be applied to all the a-type vibronic bands of the $B\,^2A_2 \leftarrow X\,^2B_1$ electronic transition. As shown in this figure, peaks of a-type vibronic transitions are expected to appear at $0.15 \pm 0.05$ Å shorter wavelengths from $T_{00}$ in diffuse clouds as DIBs.

### 3.4. *Comparison of Absorption Bands of the Thiophenoxy Radical and DIBs*

Using the simulated rotational profile of the thiophenoxy radical, we firstly compared the strongest $6a_0^2 6b_0^1$ vibronic band with the currently reported DIBs toward HD 183143 and HD 204827 (Hobbs et al. 2008, 2009) and found no agreements. As addressed in Section 3.1, the intensity of this electronic transition disperses to a number of vibronic transitions due to the structural difference between the ground and excited electronic states as described in Table A2. Thus, intensities of individual vibronic transitions are not sufficiently strong. In such cases, search of weak DIBs is worth trying by a stacking method described below.

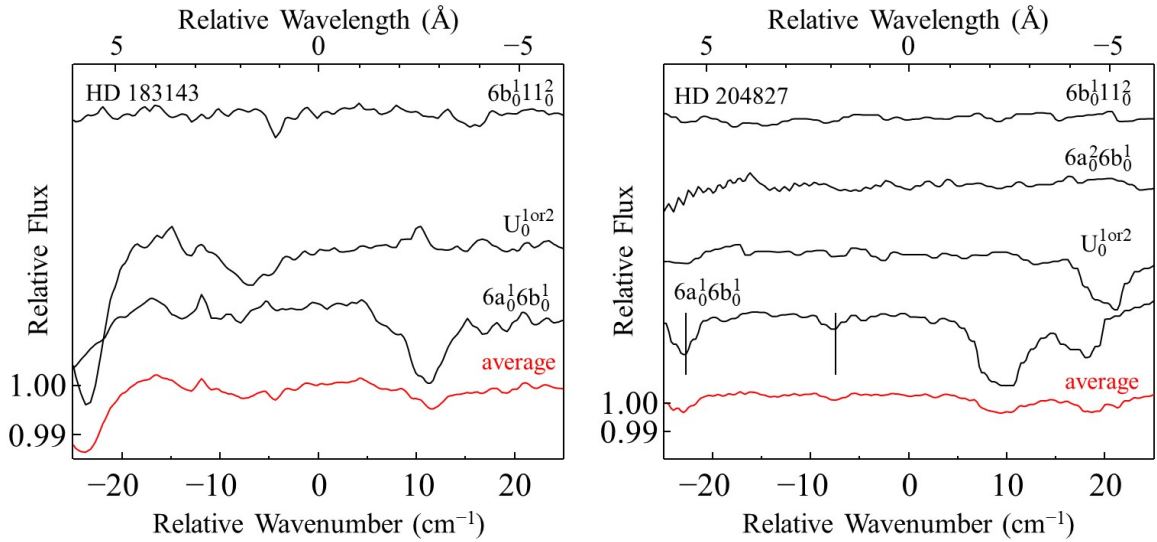

**Figure 4**. Comparisons of laboratory bands of the thiophenoxy radical with DIBs toward HD 183143 and HD 204827. DIB spectra were scanned from the reports of Hobbs et al. (2008, 2009). The horizontal axis is the wavenumber relative to the band origin of each indicated transition. The black vertical lines in the right diagram show reported DIB positions, and the other dips are features of stellar atmosphere. The red traces are averaged spectra of above black traces. Although scales of relative wavelength axis are plotted based on the band origin of the $U_0^{1or2}$ vibronic band, wavenumber dependence of the relative wavelength can be negligible in size of this figure.

Precisely determined wavelengths of the band origin in this work (Table 1) allow us to adopt the stacking technique that the reported DIBs spectra (Hobbs et al. 2008, 2009) at the positions of vibronic transitions were stacked in relative wavelengths based on the band-origin wavelengths of individual transitions as shown in Figure 4. This technique can reduce noise levels of spectra practically. By using criteria free from overlapping with stellar lines, the $6b_0^1 11_0^2$, $U_0^{1or2}$, and $6a_0^1 6b_0^1$ bands were selected for stacking. Because the $6a_0^1 6b_0^1$ band is overlapped with a stellar line, this band is excluded toward HD 183143. Toward HD 204827, the stellar line was removed from the spectrum by Gaussian fitting. Although we could not find any stacked lines, upper limits of the column densities toward HD 183143 and HD 204827 could be estimated using the procedure used by Motylewski et al. (2000). It was assumed that a signal-to-noise ratio of 5 is required to detect DIB absorption. The FWHM of the band is approximately 0.4 Å, as shown in Figure 3(b), and then the detection limits of the equivalent width in the astronomically observed spectra are 0.0015 Å toward HD 183143 (Hobbs et al.

2009) and 0.0014 Å toward HD 204827 (Hobbs et al. 2008). The oscillator strength of the $B\ ^2A_2 - X\ ^2B_1$ electronic transition was theoretically obtained to be 0.003 by TD-B3LYP/cc-pVTZ (Paper I). We suppose that the oscillation strength is not enhanced significantly by vibronic interaction. This is because the progression starting from the origin band appears weakly compared with that expected by the theoretical calculations, while the unexpected progression starting from the $6b_0^1$ forbidden transition comes out strongly (see Figure A1). Referring to the oscillator strengths of the benzyl radical by TD-B3LYP (Tonokura & Koshi 2003), we estimated an uncertainty involved in the oscillator strength to be ~30%. According to the relative intensities in Table 1, the averaged lines toward HD 183143 and HD 204827 have 6.2 and 7.4 % of this oscillator strength, respectively. These allocations then lead to ~4 × 10¹³ cm⁻² as the upper limits of the column densities for the thiophenoxy radical toward both stars. Although the upper limit reported in Paper I is approximately comparable to the present values, the assumption for intensity distribution is not appropriate as mentioned in Section 3.1. For reference, the column density of $C_{60}^+$ toward HD 183143 was reported to be 2 × 10¹³ cm⁻² (Campbell et al. 2016a).

## Summary

The vibronic bands of the $B\,^2A_2 \leftarrow X\,^2B_1$ electronic transition of the thiophenoxy radical $C_6H_5S$ produced in the hollow-cathode glow discharge was recorded using a cavity ring-down spectrometer. The precise wavelengths of the band origins may facilitate searches for absorptions of this radical in the interstellar medium. With the aid of calculations, the most prominent progression was assigned to the 6a mode (in the $B\,^2A_2$ state) starting from the $6b_0^1$ transition. The strong appearance of the 6b mode, which is originally symmetry forbidden, is attributed to the vibronic interaction between the $C\,^2B_1$ and $B\,^2A_2$ state. Although comparison of laboratory spectral data with the currently detected DIBs yielded no agreement at present, the upper limits of the column densities were estimated to be $\sim 4 \times 10^{13}$ cm$^{-2}$ both in the diffuse clouds toward HD 183143 and HD 204827.

This study was funded by Grant-in-Aid for Scientific Research on Innovative Areas (Grant No. 25108002), Grant-in-Aid for Scientific Research (C) (Grant No. 15K05395 and 18K05045), and The Mitsubishi Foundation.

Western, C. M., PGOPHER, a Program for Simulating Rotational Structure, University of Bristol, http://pgopher.chm.bris.ac.uk

### Appendix 1
#### *Calculations of Vibrational Frequencies and Franck-Condon Factors*

To assign the vibrionic structure of the thiophenoxy radical, vibrational frequencies in the upper electronic state are necessary. For the ground $X^2B_1$ and first excited $A$ $^2B_2$ states, the vibrational frequencies were theoretically calculated by Cheng et al. (2008). We calculated the structures and the frequencies of the ground and second excited $B$ $^2A_2$ states assuming $C_{2v}$ symmetry by TD-B3LYP/6-311++G(d,p) using the program package Gaussian09W (Frisch et al. 2009), as listed in Tables A1 and A2. Wilson's notations for vibrational modes were allocated based on those of $C_6H_5O$ reported by Spanget-Larsen et al. (2001).

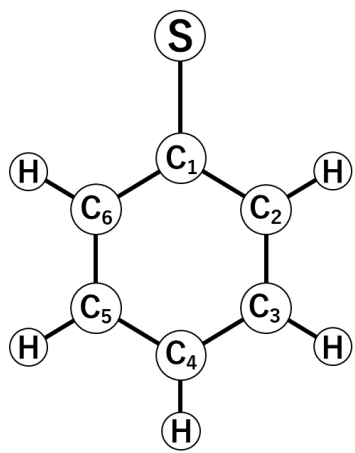

**Figure A2**. Numbering of Atoms of the thiophenoxy radical

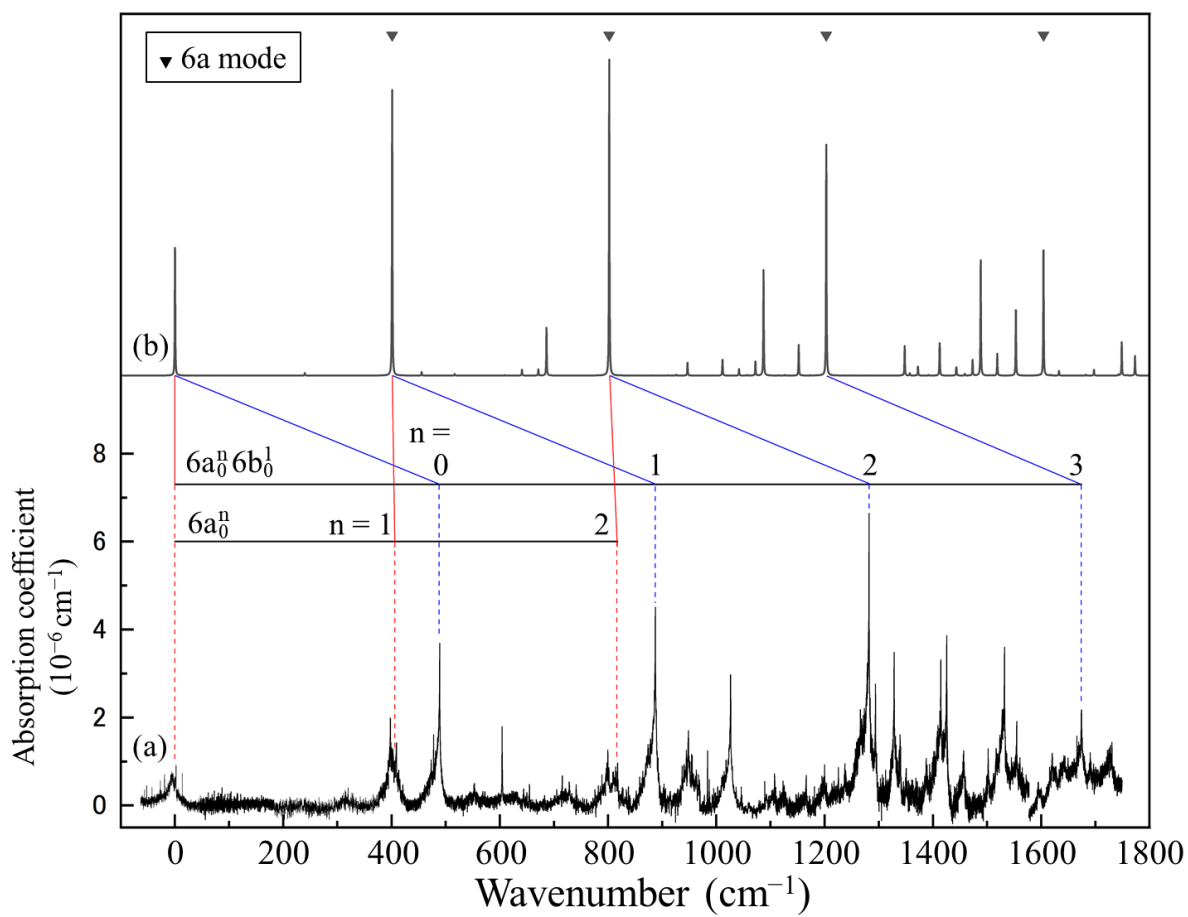

**Figure A1**. Franck-Condon calculation of the $B$ $^2A_2 \leftarrow X^2B_1$ electronic transition of the thiophenoxy radical (b). (a) The same spectrum with Figure 1(a).

To find Franck-Condon (FC) active modes in the $B\ ^2A_2 \leftarrow X\ ^2B_1$ electronic transition, FC simulation utilizing the above calculated data without consideration of vibronic interaction was performed using the FC-LabWin program package (Pugliesi et al. 2006, Schriever et al. 2009) as shown in Figure A1. The vibronic bands involving the 6b mode do not appear, since the vibronic interaction of $B\ ^2A_2$ with $C\ ^2B_1$ is not included in the present calculations. However, the prominent progression of the 6a mode is derived, suggesting that the 6a mode is the most FC active.

**Table 1**
Band Origins and Vibrational Frequencies
in the $B\,^2A_2 - X\,^2B_1$ electronic transition of the Thiophenoxy Radical

| Wavelength [a,b] $T_{00}$ (Å) | Wavenumber (cm$^{-1}$) | Vibrational Frequency (cm$^{-1}$) | Relative Intensity [c] (%) | Assignment [d] | Transition type |
|---|---|---|---|---|---|
| 5172.48 | 19327.7 | 0.0 | 1.5 | $0_0^0$ | b |
| 5066.20 | 19733.2 | 405.5 | 2.4 | $6a_0^1$ | b |
| 5045.18 | 19815.4 | 487.7 | 6.2 | $6b_0^1$ | a |
| 4962.97 | 20143.6 | 815.9 | 1.6 | $6a_0^2$ | b |
| 4945.57 | 20214.5 | 886.8 | 7.6 | $6a_0^1\,6b_0^1$ | a |
| 4911.80 | 20353.4 | 1025.7 | 5.0 | $U_0^{1\text{or}2\,e}$ | a |
| 4850.91 | 20608.9 | 1281.2 | 11.2 | $6a_0^2\,6b_0^1$ | a |
| 4839.98 | 20655.5 | 1327.8 | 5.9 | $6b_0^1\,11_0^{2\,f}$ | a |
| 4819.85 | 20741.7 | 1414.0 | 5.6 | $6b_0^1\,18a_0^{1\,f}$ | a |
| 4817.39 | 20752.3 | 1424.6 | 6.5 | $6a_0^1\,U_0^{1\text{or}2\,e}$ | a |
| 4792.70 | 20859.2 | 1531.5 | 6.1 | $6b_0^1\,9a_0^{1\,f}$ | a |
| 4760.22 | 21001.5 | 1673.8 | 3.7 | $6a_0^3\,6b_0^1$ | a |

[a] in Air

[b] Band origins corresponding to $T_{00}$ (cm$^{-1}$) in Table 2, not peak wavelengths of individual bands. Peaks of the a-type transitions appear at $0.15 \pm 0.05$ Å shorter wavelengths from $T_{00}$ in diffuse clouds as DIBs, as mentioned in Section 3.3.

[c] Observed intensities indexed using a sum of all sticks from 0 to 1750 cm$^{-1}$ in Figure 1(b) as a base of 100. Intensities of lines outside the observed range are thought to be very weak even if they exist, because Okuyama et al. (1977) did not detect them.

[d] Wilson's notation

[e] Unknown vibrational mode(s)

[f] Tentative

**Table 2**
Molecular Constants of the Thiophenoxy Radical in cm$^{-1}$

| Electronic State | Ground $X\,^2B_1$ | Excited [a] $B\,^2A_2$ | | |
|---|---|---|---|---|
| Vibrational State | 0 [b] | 0 [c] | 6b$^1$ | 6b$^1$6a$^1$ |
| $T_{00}$ | 0 | 19327.7(3) | 19815.381(8) [d] | 20214.47(2) [d] |
| $A$ | 0.1893 | | | |
| $\Delta A$ [e] | | 0.0073(5) | 0.00594(6) | 0.00663(4) |
| $\overline{B}$ [f] | 0.04848 | | | |
| $\Delta \overline{B}$ [e,f] | | −0.0017(1) | −0.002048(6) | −0.002879(4) |
| $B - C$ | 0.01222 | 0.01048 [g] | 0.01048 [g] | 0.01048 [g] |

[a] Values in parentheses denote the uncertainties of fitting and apply to the last digit of the values.
[b] The ground-state rotational constants were determined by theoretical calculations of B3LYP/cc-pVTZ in Paper I.
[c] These constants were determined by the rotational profile simulation in Paper I.
[d] In addition to this error, a 0.1-cm$^{-1}$ uncertainty of the wavelength meter needs to be considered.
[e] $\Delta A = A' - A$ and $\Delta \overline{B'} = \overline{B'} - \overline{B}$, where the constants marked with primes are of an upper state.
[f] $\overline{B} = (B + C)/2$
[g] Theoretical calculations of TD-B3LYP/cc-pVTZ.

**Table A1**
Calculated Vibrational Frequencies of the Thiophenoxy Radical in cm$^{-1}$.

| Mode | Symmetry | $X\,^2B_1$ | $B\,^2A_2$ | Wilson's Notation |
|------|----------|------------|------------|-------------------|
| 1 | $a_1$ | 3094 | 3110 | 20a |
| 2 | | 3083 | 3093 | 13 |
| 3 | | 3063 | 3073 | 2 |
| 4 | | 1543 | 1519 | 8a |
| 5 | | 1431 | 1443 | 7a |
| 6 | | 1160 | 1152 | 19a |
| 7 | | 1042 | 1058 | 9a |
| 8 | | 1004 | 1011 | 12 |
| 9 | | 971 | 947 | 18a |
| 10 | | 706 | 686 | 1 |
| 11 | | 413 | 401 | 6a |
| 12 | $a_2$ | 958 | 953 | 17a |
| 13 | | 817 | 834 | 10a |
| 14 | | 371 | 305 | 16a |
| 15 | $b_1$ | 974 | 921 | 5 |
| 16 | | 907 | 838 | 17b |
| 17 | | 739 | 719 | 4 |
| 18 | | 661 | 397 | 11 |
| 19 | | 448 | 336 | 16b |
| 20 | | 155 | 120 | 10b |
| 21 | $b_2$ | 3091 | 3092 | 20b |
| 22 | | 3072 | 3077 | 7b |
| 23 | | 1524 | 1502 | 19b |
| 24 | | 1411 | 1373 | 8b |
| 25 | | 1297 | 1321 | 14 |
| 26 | | 1259 | 1260 | 3 |
| 27 | | 1141 | 1142 | 15 |
| 28 | | 1059 | 967 | 9b |
| 29 | | 600 | 575 | 6b |
| 30 | | 289 | 277 | 18b |

Note. Theoretical calculations were performed assuming C$_{2v}$ symmetry by TD-B3LYP/6-311++G(d,p) using the program package Gaussian09W (Frisch et al. 2009). The scaling factor of 0.967 (CCCBDB) was applied. The obtained structural parameters are listed in Table A2.

**Table A2**
Calculated band lengths (in Å) and bond angles (in degree) of the Thiophenoxy Radical.

|  | $X\,^2B_1$ | $B\,^2A_2$ |
|---|---|---|
| $r\,(C_1–S)$ | 1.727 | 1.742 |
| $r\,(C_1–C_2)$ | 1.418 | 1.407 |
| $r\,(C_2–C_3)$ | 1.386 | 1.447 |
| $r\,(C_3–C_4)$ | 1.399 | 1.385 |
| $r\,(C_2–H)$ | 1.083 | 1.084 |
| $r\,(C_3–H)$ | 1.084 | 1.085 |
| $r\,(C_4–H)$ | 1.084 | 1.081 |
| $\angle C_6–C_1–C_2$ | 118.38 | 113.51 |
| $\angle C_1–C_2–C_3$ | 120.60 | 123.28 |
| $\angle C_2–C_3–C_4$ | 120.12 | 121.75 |
| $\angle C_3–C_4–C_5$ | 120.19 | 116.42 |
| $\angle C_1–C_2–H$ | 118.55 | 118.50 |
| $\angle C_2–C_3–H$ | 119.93 | 117.93 |

Note. Numbering of atoms is described in Figure A2.